\documentclass[10pt,superscriptaddress,nofootinbib,twocolumn]{revtex4}

\usepackage{amsthm,amsmath,amssymb}
\usepackage{mathrsfs}
\usepackage{appendix}
\usepackage{amstext}
\usepackage{graphicx}
\usepackage{url}
\usepackage{float}
\usepackage{bm}
\usepackage[usenames,dvipsnames]{color}
\usepackage[colorlinks=true,citecolor=blue,urlcolor=black]{hyperref}
\usepackage{dcolumn}
\usepackage{natbib}
\usepackage{booktabs}
\usepackage{graphicx}

\begin{document}

\title{Source-independent quantum random number generator against tailored detector blinding attacks}

\author{Wen-Bo Liu}\thanks{These authors contributed equally to this work}	
\author{Yu-Shuo Lu}\thanks{These authors contributed equally to this work}
\affiliation{National Laboratory of Solid State Microstructures and School of Physics, Collaborative Innovation Center of Advanced Microstructures, Nanjing University, Nanjing 210093, China.}
\author{Yao Fu}\email{yfu@iphy.ac.cn}
\affiliation{Beijing National Laboratory for Condensed Matter Physics and Institute of Physics, Chinese Academy of Sciences, Beijing 100190, China}
\affiliation{MatricTime Digital Technology Co. Ltd., Nanjing 211899, China}
\author{Si-Cheng Huang}
\affiliation{National Laboratory of Solid State Microstructures and School of Physics, Collaborative Innovation Center of Advanced Microstructures, Nanjing University, Nanjing 210093, China.}
\affiliation{State Key Laboratory of Particle Detection and Electronics, University of Science and Technology of China, Hefei 230026, China}
\author{Ze-Jie Yin}
\affiliation{State Key Laboratory of Particle Detection and Electronics, University of Science and Technology of China, Hefei 230026, China}
\author{Kun Jiang}
\affiliation{Beijing National Laboratory for Condensed Matter Physics and Institute of Physics, Chinese Academy of Sciences, Beijing 100190, China}
\author{Hua-Lei Yin}\email{hlyin@nju.edu.cn}
\affiliation{National Laboratory of Solid State Microstructures and School of Physics, Collaborative Innovation Center of Advanced Microstructures, Nanjing University, Nanjing 210093, China.}
\author{Zeng-Bing Chen}\email{zbchen@nju.edu.cn}
\affiliation{National Laboratory of Solid State Microstructures and School of Physics, Collaborative Innovation Center of Advanced Microstructures, Nanjing University, Nanjing 210093, China.}
\affiliation{MatricTime Digital Technology Co. Ltd., Nanjing 211899, China}

\begin{abstract}
Randomness, mainly in the form of random numbers, is the fundamental prerequisite for the security of many cryptographic tasks.
Quantum randomness can be extracted even if adversaries are fully aware of the protocol and even control the randomness source.
However, an adversary can further manipulate the randomness via tailored detector blinding attacks, which are hacking attacks suffered by protocols with trusted detectors.
Here, by treating no-click events as valid events, we propose a quantum random number generation protocol that can simultaneously address source vulnerability and ferocious tailored detector blinding attacks.
The method can be extended to high-dimensional random number generation.
We experimentally demonstrate the ability of our protocol to generate random numbers for two-dimensional measurement with a generation speed of 0.1 bit per pulse.
\end{abstract}
\maketitle

\section{Introduction}
The unpredictability of random numbers was originally intended to refer to a lack of correlation between numbers.
In the current study, pseudorandom numbers~\cite{pserand,pserand3} are obtained through deterministic formulas implying some correlation of these numbers and, hence, some predictability of subsequent numbers.
For the physical true random numbers~\cite{truerand1,truerand2}, the source of its randomness has not been fully studied.
In contrast, quantum random numbers~\cite{QRNG1,QRNG2} are considered to have inherent randomness based on the completeness of quantum mechanics.
Quantum random number generators (QRNGs) have thus been widely investigated to obtain unpredictable random numbers.
In addition to their lack of correlation, the practical security of quantum random numbers has received considerable attention as their fields of application~\cite{app1,app2,app4} expand to cryptographic tasks~\cite{yin2016practical,app3,fu2015long,app5,liu2021homodyne}.

A QRNG typically consists of a randomness source and a detection device. The randomness source provides light with quantum properties, and the detection device extracts randomness by measurements of light in a superposition state.
As a solution to almost all security concerns, device-independent QRNGs~\cite{DI1,DI2,DI3,DI4} are the most stringent, making no assumption about either randomness sources or detection devices.
Recently, device-independent QRNGs that can extract random numbers after deducting the consumed randomness have been implemented for the first time.
The net gains reached 2290 bps~\cite{DI5}, 3606 bps~\cite{DI6} and 3718 bps~\cite{DI7}.
However, they all required approximately 10 hours to accumulate data, which would lead to high latency in practical use.
Furthermore, random numbers are consumed rapidly in most cryptographic tasks.
Thus, we unavoidably consider the trade-off between security and the generation rate~\cite{lowlatency,SDIF,MDIF,QSI}.
An adoptable choice is the source-independent QRNG (SI-QRNG)~\cite{SI1,SDI1,SDI2,SDI3,cheng2022mutually}, in which the detection devices are assumed to be trusted by well characterizing them.
There is no secure assumption on the randomness source and the channel between the source and the detection device. Different from device-independent QRNGs, SI-QRNGs can measure both discrete variables~\cite{SI1} and continuous variables~\cite{SDI1,SDI2,SDI3,cheng2022mutually}. Here we focus on the discrete-variable QRNG since it needs no additional local oscillators and is realized by a single measurement. In practice, SI-QRNG has a wide selection of untrusted sources, from lasers to light bulbs to sunlight, depending on the situations, thus becoming a popular choice.

Perfectly characterizing detectors is complex and difficult~\cite{afterpulse,sunlight}.
Researchers have tried to solve the known vulnerabilities one by one.
But they believe in the assumption that detectors can detect a single photon under any attack.
Tailored detector blinding attacks~\cite{APDattack,wiechers2011after,lydersen2011superlinear}, first introduced in quantum key distribution~\cite{app6}, is the most powerful attack targeting detectors.
It causes the detector to respond to signals up to a certain intensity by change the physical state of detectors.
The adversary Eve thus can manipulate the detector using trigger light with specific optical power and determines the detection outcomes with a probability of almost 100\%.
Such attacks can be launched on either avalanche photodiodes~\cite{APDattackfull} or superconducting nanowire single-photon detectors (SNSPDs)~\cite{nanoattack}.

Inspired by the interpretation of no-click events in Bell tests~\cite{brunner2014bell}, we find that the change of the physical state of detectors breaks the fair sampling assumption.
In this work, we present a source-independent protocol that is secure against the tailored detector blinding attacks by counting no-click events.
Additionally, our protocol has composable security against quantum coherent attacks and can be easily expanded to high-dimensional measurement cases.
We experimentally demonstrate the feasibility of generating random numbers in the two-dimensional measurement case.
Detector imperfections such as dark count and after pulse are also considered.
Our protocol achieves higher security than previous SI-QRNGs and maintains a meaningful generation rate.
In our experiments, we realize the generation rate 0.103 with 1 Gb of data accumulation.
For low-latency applications, our experimental system is able to generate 640 kbit quantum random numbers every 2 seconds with a 5 MHz experimental system.
The extracted quantum random numbers pass the NIST test.

\section{Tailored detector blinding attacks}
Detector blinding attacks originate from flaws in the single-photon detector.
Strictly speaking, only in specific mode can detectors detect a single photon.
After changing conditions such as bias voltage and temperature, the detector may require stronger light to respond.
This flaw gives Eve the opportunity to change conditions by injecting special light, and then arbitrarily set the threshold of detectors in his favor, which is the tailored detector blinding attack~\cite{APDattack}.
We first construct a threshold detection model for detectors under bright illumination.
Based on this model, we describe the attacks we aim to solve.
Finally, we describe the performance of attacks in high-dimensional measurement cases.
\begin{figure*}[tb]
\centering
\includegraphics[width=17.2cm]{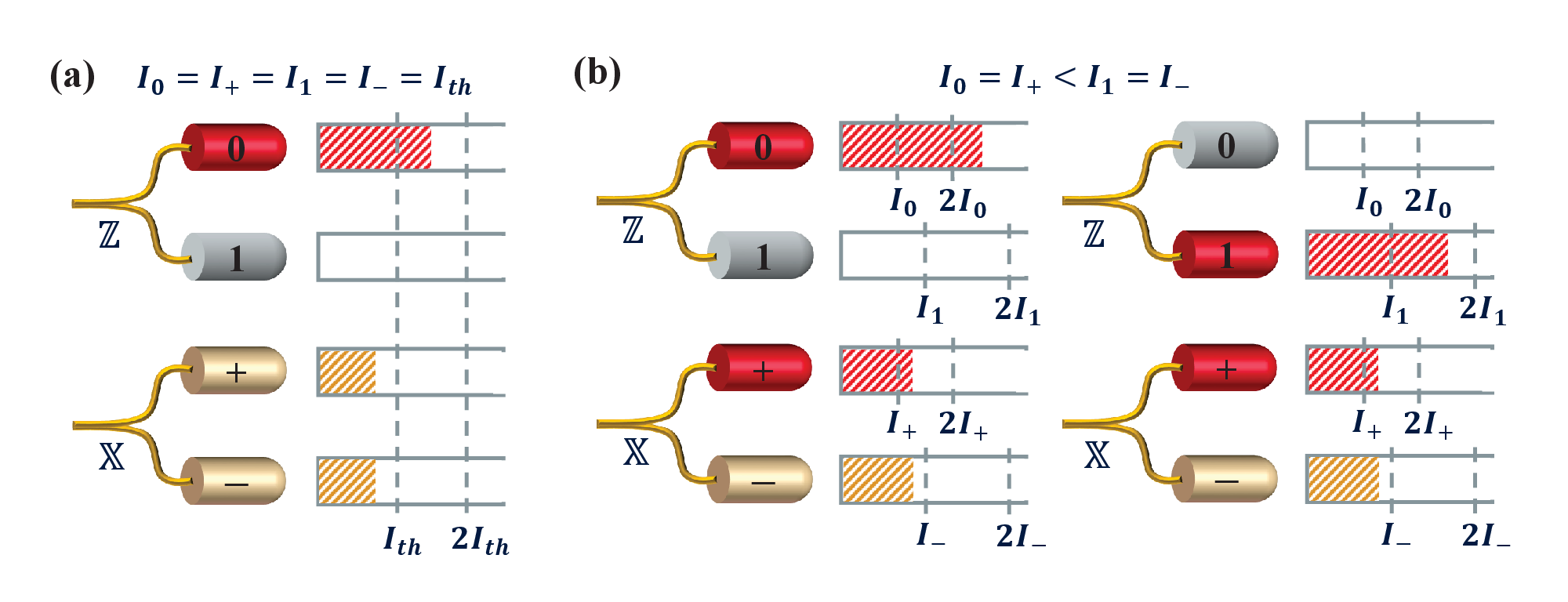}
\caption{Tailored detector blinding attacks in two-dimensional measurement. (a) The case in which both detectors have the same thresholds. Although Eve controls the outcomes of measuring signals in $\mathbb{Z}$, both detectors do not fire if Alice happens to measure signals in $\mathbb{X}$. (b) The case in which the detector representing $\vert +\rangle$ has a lower threshold than the other. When Eve controls the outcomes of measuring signals in $\mathbb{Z}$, she can also cause the detector representing $\vert +\rangle$ to fire if Alice happens to measure signals in $\mathbb{X}$.}
\label{fig1}
\end{figure*}

\subsection{Threshold detection model}
As the receiver, Alice detects signals randomly in one of two incompatible bases $\mathbb{X}$ and $\mathbb{Z}$.
Without loss of generality, we agree that the outcomes in $\mathbb{Z}$ are used to generate raw random numbers and the outcomes in $\mathbb{X}$ are used to judge the amount of information obtained by Eve.
In the two-dimensional measurement scheme, we notate the eigenstates of $\mathbb{Z}$ as $\{\left| 0 \right\rangle, \left| 1 \right\rangle\}$, and the eigenstates of $\mathbb{X}$ as $\{\left| \pm \right\rangle = \frac{1}{\sqrt{2}}(\left| 0 \right\rangle \pm \left| 1 \right\rangle)\}$.
When the $\mathbb{X}$ basis is chosen, the outcome $\left| + \right\rangle$ is considered the correct outcome, and the outcome $\left| - \right\rangle$ is an error event~\cite{SI1}.

We define the threshold of a detector as the intensity $I$, which means that the detector fires when the intensity of the signal is stronger than $I$ and not when it is equal to or weaker than $I$.
In the tailored detector blinding attack scenario, Eve can arbitrarily determine the value of $I$ by exploiting the tailored bright illumination, and Alice cannot obtain this value unless additional monitoring is performed.
Under the active-basis-choice, we can assume that the threshold of the detector representing $\left| 0 \right\rangle$ and $\left| + \right\rangle$ is $I_0=I_+$ and that the threshold of the detector representing $\left| 1 \right\rangle$ and $\left| - \right\rangle$ is $I_1=I_-$.
$I_0 = I_1 = 0$ when detectors are in the single-photon response mode.
When Eve sends signals with bright illumination, the thresholds of the different detectors are governed by Eve.
Here, we assume that the detectors have perfect efficiency.
The inefficiency occurs only when $I_0 = I_1 = 0$ and the detector can be considered a perfect detector with some loss in the channel.
When the thresholds of the detectors are higher than $0$, the physical property of the detectors is changed.
The signal is detected in the form of light intensity, and there is no concept of detection efficiency.

\subsection{Attack description}
We first state that the Eve's control over the threshold is not instantaneous.
The attack we discuss here does not allow Eve to change the threshold of the detector every detection window because Eve blinds the detector through bright continuous-wave.
This assumption is realistic and avoids an ideal attack: sending  $\left| + \right\rangle$ all the time, but changing the thresholds of detectors representing $\left| 0 \right\rangle$ and $\left| 1 \right\rangle$ to determine which detector responds each time.
Second, we assume that Eve is greedy, and she only wants the value she chooses to be detected, not a value that is more likely to be detected.
In this regard, Eve's method changes the detector threshold so that the signal he sends accurately enters a certain detector, and the response he expects occurs.

A simple attack for Eve is to tune the detectors to have the same threshold $I_{th}$, as shown in Fig.~\ref{fig1}a.
Eve wants Alice to obtain an outcome specified by Eve when Alice measures the signal in $\mathbb{Z}$.
In other words, a signal with intensity $I_e  > I_{th}$ enters either the detector representing $\left| 0 \right\rangle$ or the detector representing $\left| 1 \right\rangle$ in accordance with Eve's arrangement.
At the same time, Eve requires the detector representing $\left| - \right\rangle$ not to fire if Alice happens to measure the signal in $\mathbb{X}$.
Since half of the photons in the signal arrive at the $\left| + \right\rangle$ detector and the others arrive at the $\left| - \right\rangle$ detector, Eve sets $0.5 I_e \leqslant I_{th}$ to cause a no-click event.
In squashing models~\cite{Lusquash1,Losquash,Lusquash2,Kisquash}, no-click events are treated as receiving vacua and thus are discarded without increasing the error count.
Therefore, by emitting signals with $I_{th}<I_e \leqslant 2I_{th}$, Eve can control the outcomes of $\mathbb{Z}$-basis measurements without increasing the error rate in $\mathbb{X}$.

The general case is that the thresholds of the different detectors are different, as shown in Fig.~\ref{fig1}b,
A more favorable option for Eve is $I_+ < I_-$ since $\left| + \right\rangle$ represents the correct outcome.
For the active-basis-choice, we have $I_0=I_+$ and $I_1=I_-$, which means that $I_0 < I_1$.
In this case, Eve can cheat both bases at the same time, i.e., she controls the outcomes in $\mathbb{Z}$ while ensuring that only the $\left| + \right\rangle$ detector fires in $\mathbb{X}$.
If Eve wants Alice to obtain an outcome of $\left| 0 \right\rangle$, she emits a signal with $I_e>I_0$, and all photons in it are sent to the $\left| 0 \right\rangle$ detector under the $\mathbb{Z}$ basis.
If Eve wants Alice to obtain an outcome of $\left| 1 \right\rangle$, she emits the signal with $I_e>I_1$, and all photons in it are sent to the $\left| 1 \right\rangle$ detector under the $\mathbb{Z}$ basis.
To make the outcomes in $\mathbb{Z}$ credible, she also requires $0.5I_e \leqslant I_-$ and $0.5I_e > I_+$ under the $\mathbb{X}$ basis.
Overall, the intensity of the signal should be $\max\{I_1,2I_0\} < I_e \leqslant 2I_1$, which does not violate the premise $I_0 < I_1$.

\subsection{d-dimensional case}
Tailored detector blinding attacks also work in the $d$-dimensional measurement scenario.
Two measurement bases $\mathbb{X}$ and $\mathbb{Z}$ are both $d$-dimensional and ideally have the relation $\left\vert _{z}\langle i\vert j \rangle_x\right\vert = 1/\sqrt{d}$ between any eigenstate $\vert i \rangle_z$ ($i \in \{1,2,...,d\}$) of $\mathbb{Z}$ and any eigenstate $\vert j \rangle_x$ ($j \in \{1,2,...,d\}$) of $\mathbb{X}$.
The outcome $\left| 0 \right\rangle_x$ is the correct outcome in $\mathbb{X}$.
Eve will emit signals with intensity $I_{th}<I_e \leqslant d I_{th}$ if she sets the same threshold $I_{th}$ for all detectors.
When Alice measures the $\mathbb{X}$ basis, the light intensity entering each detector is $I_e/d$, which is less than the threshold $I_{th}$.

The situation will be slightly more complicated if Eve wants to control both bases perfectly.
She can adjust the threshold of the detector representing $\left| 0 \right\rangle_x$ to the lowest among all detectors' thresholds.
Thus the $\left| 0 \right\rangle_x$ detector is the one that is most easily responded when using the $\mathbb{X}$ basis to measure signals that are the eigenstate in the $\mathbb{Z}$ basis.
The light intensity should be $d$ times higher than the threshold of the $\left| 0 \right\rangle_x$ detector to ensure the response of the detector.
To avoid multiple-click events in $\mathbb{X}$, the light intensity should also be less than $d$ times the sub-smallest threshold.
This, in turn, constrains the thresholds of the other detectors to be less than $d$ times the sub-smallest threshold.
Otherwise, those detectors with a threshold higher than $d$ times the sub-smallest threshold will fail to fire because the light intensity is not sufficient.

\section{Defensive strategy}
In general, Eve controls the detectors while causing no click in the $\mathbb{X}$ basis, which is a hint for us.
In terms of this hint, we should reconsider what no click means.
First, we briefly review the concept of squashing models and analyze why this hint has been ignored in previous works.
Then, we introduce a strategy for handling this hint, which modifies previous squashing models.
The uncertainty relation for smooth entropy is used as a critical tool for generating quantum random numbers that are secure against general attacks.
Finally, we generalize the security analysis to the $d$-dimensional case.
\begin{figure}[t]
\centering
\includegraphics[width=8.6cm]{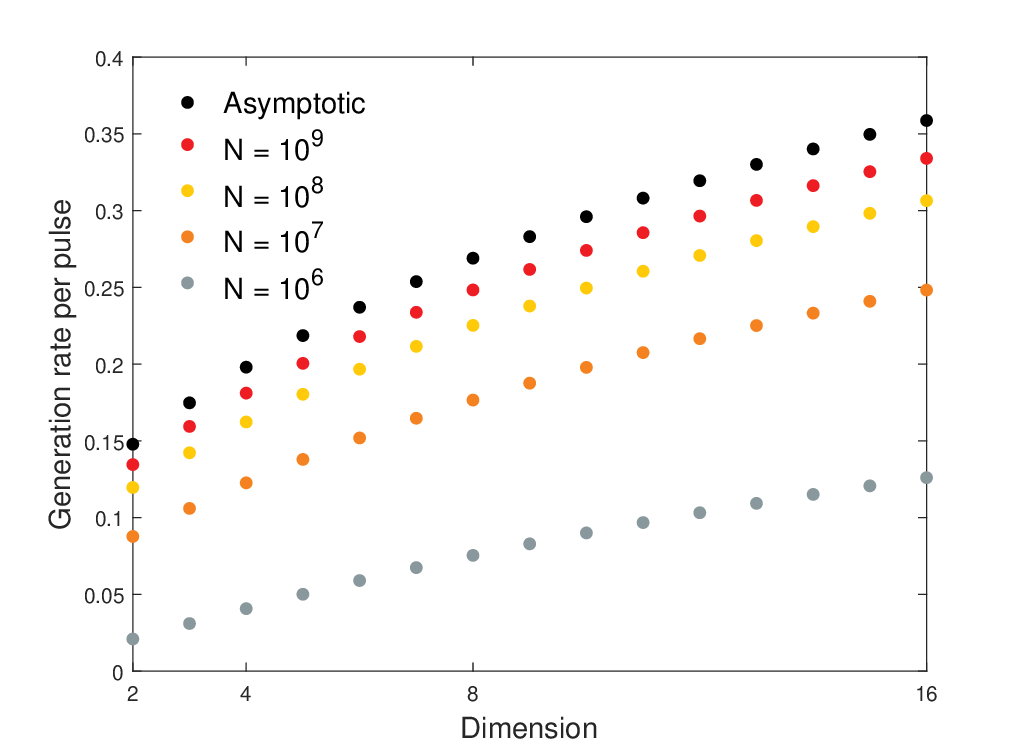}
\caption{Dimensions of the measurement basis and generation rates per pulse. The asymptotic case is investigated, in which the amount of data is infinite. Scatter points in different colors represent cases in which the numbers of emitted pulses $N$ are $10^9$, $10^8$, $10^7$ and $10^6$.}
\label{fig2}
\end{figure}

\subsection{Squashing model}
The dimension of the signals output from the channel is unknown since the channel is controlled by Eve.
However, security analysis is usually qubit-based for two-dimensional measurements by virtue of simplicity.
The squashing model~\cite{Lusquash1,Losquash,Lusquash2,Kisquash} is developed to resolve this conflict.
A squashing operation is applied to the signal, which virtually maps the multi-photon signal into a qubit.
A virtual qubit measurement on this virtual qubit follows.
Therefore, qubit-based security analysis is applicable for sources with unknown dimensions.

Measuring a qubit yields one of two outcomes corresponding to its two eigenstates.
However, an unknown signal subjected to two-dimensional measurement actually yields one of four outcomes: a single click in one detector, a single click in the other detector, a double-click or no click.
To reconcile this difference in outcomes, there are three treatments for different outcomes of signals.
Single-click events in either detector are naturally related to the outcomes of measuring qubits.
Double-click events are valid events but tell us nothing about randomness.
They are used to evaluate the upper bound of the error rate~\cite{Losquash}.
Note that another squashing model~\cite{Lusquash1,Lusquash2} randomly assigns values for double-click events, and thus has a lower error rate and higher randomness consumption.

No-click events are regarded as vacua after losses.
The positions of the losses in both bases are assumed to be uniformly random.
Under this assumption, there are no qualms about discarding no-click events without disturbing the error rate.
The protocol treating no-click events as vacua is described in ''Supplemental Document''.
However, tailored detector blinding attacks break this confidence since the thresholds of the detectors can be changed such that a signal can definitely cause clicks in one basis and no click in the other.
In the worst case, all no-click events in $\mathbb{X}$ are caused by tailored detector blinding attacks.
Therefore, squashing models fail under such attacks.

\subsection{Security analysis}
The key point of our security analysis is how to securely deal with no-click events.
Tasks such as Bell tests and device-independent quantum key distribution also suffer from the loophole introduced by no-click events, called the fair sampling loophole.
An ingenious method is presented in Bell tests~\cite{brunner2014bell}, in which some no-click events are retained to close this loophole; otherwise, the experimental results may have been screened by unknown factors.
Inspired by this idea, we retain all no-click events.
No-click events should have the same status as double-click events since they both have no randomness and can cover up attacks.
Therefore, we treat no-click events in the same way as double-click events.
They are error events in the $\mathbb{X}$ basis and correct events in the $\mathbb{Z}$ basis.
The squashing model can now work under tailored detector blinding attacks.

Furthermore, considering $I_- > I_+ > 0$, it seems that no no-click event exists.
In response to this situation, the $\left| + \right\rangle$ detector should be randomly assigned by Alice.
Eve thus cannot accurately forecast it and has at most a 50\% chance of firing in the $\left| + \right\rangle$ detector.
Since we need only a small percentage of rounds to measure $\mathbb{X}$, the consumption of random numbers for deciding which detector will be used to measure $\left| + \right\rangle$ in each round is not an unbearable burden.

Our security analysis adopts the uncertainty relation for smooth entropy~\cite{Smooth,QKDex} to offer security against the most general attacks.
This relation involves three parties, namely, the user Alice, the virtual user Bob and the adversary Eve, and is expressed as
\begin{equation}\label{unc}
\begin{aligned}  
H_{\rm{min}}^{\epsilon}(\textbf{Z}_{\rm A} \vert \textbf{E}) + H_{\rm{max}}^{\epsilon}(\textbf{X}_{\rm A} \vert \textbf{B}) & \geqslant q,
\end{aligned}
\end{equation}
where $\textbf{X}_{\rm A}$ ($\textbf{Z}_{\rm A}$) means that Alice measures her system ${\rm A}$ in the $\mathbb{X}$ ($\mathbb{Z}$) basis.
The bound $q$ is an evaluation of the ``incompatibility" of the measurement bases $\mathbb{X}$ and $\mathbb{Z}$.
The smooth min-entropy $H_{\rm{min}}^{\epsilon}(\textbf{Z}_{\rm A} \vert \textbf{E})$ is Eve's minimum uncertainty about $\textbf{Z}_{\rm A}$, which quantifies how much randomness can be extracted.
The smooth max-entropy $H_{\rm{max}}^{\epsilon}(\textbf{X}_{\rm A} \vert \textbf{B})$ is related to the error rate of Bob guessing the value of $\textbf{X}_{\rm A}$.
Bob is introduced as a virtual trusted user.
He works with Alice and guesses the result of measuring the signal in the $\mathbb{X}$ basis.
Ideally, measuring the signal in the $\mathbb{X}$ basis leads to $\textbf{X}_{\rm A} = \left| + \right\rangle$.
Bob thus can guess $\textbf{X}_{\rm A} = \left| + \right\rangle$ to obtain a higher random number generation rate if Eve abandons her attack.

\subsection{d-dimensional case}
We can extend the security analysis against tailored detector blinding attacks to the $d$-dimensional measurement scenario.
In $d$-dimensional measurement, the squashing model will squash the input signal into a qudit.
There are $d$ possible outcomes when one measures the qudit in any qudit basis.
The possible real outcomes of signals are no-click events, multiple-click events, and $d$ kinds of single-click events.
The first two types of events are considered error events in $\mathbb{X}$-basis measurement and correct events in $\mathbb{Z}$-basis measurement.
Single-click events are naturally related to the qudit measurement outcomes.
Similarly, the $\left| 0 \right\rangle_x$ detector must be randomly selected.

\section{Protocol description}
Because the protocol is source independent, it focuses only on the measurement of unknown light and subsequent processing steps. Nevertheless, we offer a state preparation step before measurement, considering that Alice can provide an untrusted source to generate favorable signals and then improve the generation rate if Eve does not attack.
We directly describe our protocol in the $d$-dimensional measurement case.
Alice measures the signals in two partially complementary bases $\mathbb{X}$ and $\mathbb{Z}$ with eigenstates $\{\left| i \right\rangle_x\}$ and $\{\left| j \right\rangle_z\}$ ($i,j\in \{0,1,...,d\}$), respectively.
Here, $d$ is the measurement dimension.

\emph{State preparation.} According to the specific structure of the detection devices, the source is expected to emit $N$ signals that cause only the $\left| 0 \right\rangle_x$ detector to fire.
Although the source is not trusted, Bob can guess that the outcomes of measuring signals in the $\mathbb{X}$ basis are always $\left| 0 \right\rangle_x$.
This may help improve the extractable randomness in practice.
This step is public. Eve can change or replace signals at will before they enter the detection device.

\emph{$d$-dimensional measurement.} Alice partially trusts her detection equipment.
She randomly measures signals in basis $\mathbb{X}$ or $\mathbb{Z}$ with probability $p_x$ or $p_z=1-p_x$, respectively.
Usually, $p_x$ is much lower than $p_z$, which is beneficial for the generation rate.
When measuring signals in $\mathbb{X}$, she should randomly choose one of the detectors to detect $\left| 0 \right\rangle_x$.

\emph{Post-processing.} In the $\mathbb{X}$ basis, the measurement outcomes can be divided into two parts: $N^c_x$ and $N^e_x$.
$N^c_x$ is the number of correct outcomes in which only the detector that measures $\left| 0 \right\rangle_x$ fires.
Other outcomes, including multiple-click events, single-click events on the incorrect detector and no-click events, are considered error outcomes and are counted in $N^e_x$.
In the $\mathbb{Z}$ basis, we care only about single-click events, the total number of which is $N^s_z$.

\emph{Extract randomness.} We analyze randomness $H_{\rm{min}}^{\epsilon}(\textbf{Z}_{\rm A} \vert \textbf{E})$ we can extract by the uncertainty relation for smooth entropy in Eq.~(\ref{unc}).
To bound $H_{\rm{max}}^{\epsilon}(\textbf{X}_{\rm A} \vert \textbf{B})$, we should evaluate the conflict between the guesses of Bob and the measurement outcomes of Alice on the $\mathbb{X}$ basis.
This entropy formula concerns the outcomes that we suppose to use the $\mathbb{X}$ basis to measure signals that have actually been measured in $\mathbb{Z}$.
Although we cannot obtain the outcomes directly, we can evaluate the probability that Bob guessed incorrectly by randomly choosing several rounds to test the outcome distribution in $\mathbb{X}$.
This is why we introduce the monitoring basis $\mathbb{X}$, and the bit error rate $e_x=N^e_x/N_x$ reflects the probability that Bob guessed incorrectly in the asymptotic regime.

When considering the finite-key effect, we can apply the random sampling method to $e_x$ and obtain the upper bound $\bar{e}_x=e_x+\gamma(N_z,N_x,e_x,\epsilon_{rand})$ in the signals measured in $\mathbb{Z}$ with failure probability $\epsilon_{rand}$, where $\gamma$ is a fluctuation that can be numerically determined~\cite{randsamp}:
\begin{equation}
\gamma(n, k, \lambda, \epsilon)=\frac{\frac{(1-2\lambda)AG}{n+k}+
\sqrt{\frac{A^2G^2}{(n+k)^2}+4\lambda(1-\lambda)G}}{2+2\frac{A^2G}{(n+k)^2}},
\end{equation}
with $0 < \lambda < \lambda + \gamma \leq 0.5 $, $A=\max\{n,k\}$ and $G=\frac{n+k}{nk}\ln{\frac{n+k}{2\pi nk\lambda(1-\lambda)\epsilon^{2}}}$.

Furthermore, only the single-click events in $\mathbb{Z}$ are valid random numbers.
Other events, such as multiple-click events and no-click events, have no extractable randomness.
The upper bound of the error rate in these single-click rounds~\cite{Losquash} is $\bar{\phi}_z=(\bar{e}_x\times N_z)/N^s_z$, which means that all errors occurred in single-click events.

The smooth entropy $H_{\rm{max}}^{\epsilon}(\textbf{X}_{\rm A} \vert \textbf{B})$ is therefore limited by $h_d(\bar{\phi}_z)$, where $h_d(x) = -x\log_{2}(x/(d-1))-(1-x)\log_{2}(1-x)$ is the Shannon entropy function~\cite{Shan1,Shan2} in the case of $d$-dimensionality.
The entropy $h_d(x)$ is concave and reaches its maximum value of $\log_{2}d$ at $x=(d-1)/d$.
When $x$ is greater than $(d-1)/d$, i.e., the error rate is higher than that of random guesses, $H_{\rm{max}}^{\epsilon}(\textbf{X}_{\rm A} \vert \textbf{B})$ is set to $\log_{2}d$. In two-dimensional measurement, $H_{\rm{max}}^{\epsilon}(\textbf{X}_{\rm A} \vert \textbf{B})$ reduces to the binary Shannon entropy function $h_2(x)=-x\log_{2}(x)-(1-x)\log_{2}(1-x)$ with error rate $x$.
The entropy $h_2(x)$ reaches its maximum value of $1$ at $e_x=0.5$.
When $x$ is greater than $0.5$, we set $H_{\rm{max}}^{\epsilon}(\textbf{X}_{\rm A} \vert \textbf{B}) = 1$.

Additionally, we need the leftover hashing method~\cite{leftover} to distill random numbers from the randomness $H_{\rm{min}}^{\epsilon}(\textbf{Z}_{\rm A} \vert \textbf{E})$.
For random number generation tasks, we focus on the secrecy in the composable security.
In our protocol, there are three components that contribute to secrecy: smooth entropy, random sampling fluctuation and leftover hashing.
They all have probabilities of failure.
The failure probabilities of these components are labeled $\epsilon$, $\epsilon_{rand}$ and $\epsilon_{hash}$, respectively.
According to the composable security, the protocol has $\varepsilon_{\rm sec}$-secrecy when $\varepsilon_{\rm sec}\geqslant \epsilon+\epsilon_{rand}+\epsilon_{hash}$.
For simplicity, we take $\epsilon=\epsilon_{rand}=\epsilon_{hash}=\varepsilon_{\rm sec}/3$.
Through leftover hashing, we can generate a random number string of length $\ell$:
\begin{equation}
\begin{aligned}
&\frac{1}{2}\sqrt{2^{\ell - H_{\rm{min}}^{\epsilon}(\textbf{Z}_A\vert \textbf{E})}} \leqslant \epsilon_{hash},\\
&\ell \geqslant H_{\rm{min}}^{\epsilon}(\textbf{Z}_A\vert \textbf{E}) - 2\log_{2}\frac{1}{2\epsilon_{hash}}.
\end{aligned}
\end{equation}
In accordance with Eq.~(\ref{unc}), we finally obtain the length of secret random numbers with $\varepsilon_{\rm sec}$-secrecy is given by
\begin{equation}\label{len}
\begin{aligned}  
\ell & \geqslant N_{z}^{s}[q-h_d(\bar{\phi}_z)]-2\log_2\frac{3}{2\varepsilon_{\rm sec}}-n_{seeds},
\end{aligned}
\end{equation}
where $\bar{\phi_z}$ is the upper bound of the error rate assuming that we used the $\mathbb{X}$ basis to measure the signals leading to single-click events in the $\mathbb{Z}$ basis.

For the active-basis-choice, we need to consume some random numbers while generating them.
First, the basis choice consumes approximately $N_x \log_{2}N$~\cite{SI1}.
Second, we should assign the detection channel for measuring the eigenstate $\vert +\rangle$ every time we measure the state in $\mathbb{X}$.
This consumes approximately $N_x \log_{2}d$, where $d$ is the dimensionality of the measurement.
Therefore, the term $n_{seeds}$ in Eqs.~(\ref{len}) is $n_{seeds}=N_x \log_{2}N+N_x \log_{2}d$.

The relation between the extracted randomness per pulse and the dimensionality of the measurement basis is shown in Fig.~\ref{fig2}.
For simplicity, we assume perfect detection here with a dark count of $10^{-5}$.
In the simulation, we assume that the states are coherent states.
The yield when the signal contains $n$ photons and the measurement in $\mathbb{X}$ causes a single click on the $\vert 0 \rangle_x$ detector is $Y_{n}^{\vert 0 \rangle_x} =(1-p_d)^{d-1}- (1-p_d)^{d}(1-\eta)^n$, where $d$ is the dimensionality of measurement, $p_d$ is the dark count and $\eta$ evaluates the total loss, including the detection inefficiency.
The gain of this kind of single-click event is $Q_{\mu}^{x} =(1-p_d)^{d-1}-(1-p_d)^{d}e^{-\mu'}$.
Here, we can consider the light intensity and loss collectively as $\mu' = \mu\eta$, since both of them are insecure.
The experiment indicates that the misalignment error is $e_d = 0.004$.
We roughly use $N^e_x = N_x(1-Q_{\mu}^{x}+e_d Q_{\mu}^{x})$.

The yield when the signal contains $n$ photons and the measurement in $\mathbb{Z}$ causes a single click on one detector is $Y_{n}^{sc}=(1-p_d)^{d-1}\left[(1-(d-1)\eta/d)^n - (1-\eta)^n (1-p_d)\right]$.
The gain of all single click events is $Q_{\mu}^{z} =d(1-p_d)^{d-1}e^{-(d-1)\mu'/d}-d(1-p_d)^{d}e^{-\mu'}$.
We have $N^s_z = N_z Q_{\mu}^{z}$.
We optimize both the light intensity and the basis choice ratio.
Although the consumption of random seeds increases as the dimension increases, the increase in dimension is beneficial for the extracted randomness per pulse.
Different data sizes have an impact on the generation rate.
Note that the data size here refers to the number of pulses sent.
Even if the data size is only $10^6$, the random number can be extracted effectively, which implies the possibility of real-time random number generation.

\begin{figure*}[t]
\centering
\includegraphics[width=17.2cm]{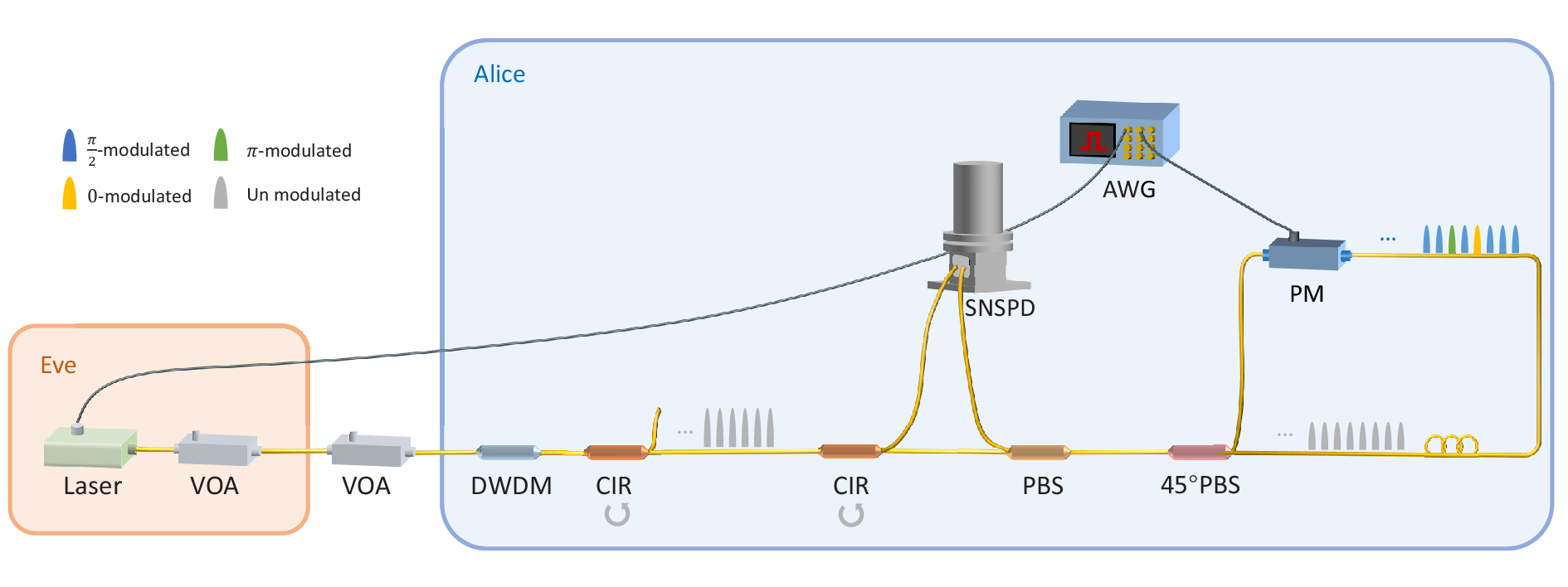}
\caption{Experimental setup. The untrusted source is composed of a laser and a variable optical attenuator (VOA). The partially trusted detection equipment includes a dense wavelength division multiplexer (DWDM), two circulators (CIR), a polarization beam splitter (PBS), a polarization beam splitter with $45 ^\circ$ alignment ($45^\circ$ PBS), a phase modulator (PM) controlled by an arbitrary waveform generator (AWG), and a superconducting nanowire single-photon detector (SNSPD). All optical fibers are polarization-maintaining fibers.}
\label{fig_Experimentsetup}
\end{figure*}

\section{Experimental implementation}
We experimentally implement our QRNG protocol using the setup shown in Fig.~\ref{fig_Experimentsetup}, which includes an untrusted randomness source and a trusted detection device with the structure disclosed. Random number generation with two-dimensional measurement is demonstrated.
The measurement bases used here are the polarization bases, and all fiber paths in our setup are polarization-maintaining fibers.
We refer to the state that propagates through the slow (fast) axis of the polarization-maintaining fiber as the eigenstate $\vert H \rangle$ ($\vert V \rangle$) of basis $\mathbb{Z}$.

In the detection part, a dense wavelength division multiplexer and a circulator are utilized to resist wavelength-dependent attacks~\cite{li2011attacking} and detector backflash attacks~\cite{pinheiro2018eavesdropping}, respectively. The DWDM can be replaced with a DWDM series to better isolate other wavelengths.
The optical pulses from the source enter the circulator and are fully transmitted through a polarization beam splitter (PBS).
The pulses are split by a $45^\circ$-aligned polarization beam splitter ($45^\circ$ PBS) and enter a Sagnac interferometer.
In the Sagnac interferometer, a phase modulator (PM) driven by an arbitrary waveform generator is utilized to realize the active-basis-choice by modulating the relative phase between clockwise and anticlockwise propagating pulses.
The anticlockwise propagating pulses arrive at the PM with a 25 ns delay relative to clockwise propagating pulses, although they pass through the same fiber.
The selections of the measurement basis and the detector representing $\vert + \rangle$ are commanded by quantum random numbers generated from a previous quantum key distribution experiment~\cite{yin2020experimental}.
In the experiment, the sequence with length $10^4$ is circularly fed to the AWG.
The probability of selecting the $\mathbb{Z}$ basis is $99.95\%$. When the $\mathbb{Z}$ basis is chosen for measurement, the PM adds a $\pi/2$ phase shift on the earlier arrived pulse. When the $\mathbb{X}$ basis is chosen, to avoid the attack with $I_0 \neq I_1$, PM randomly adds a $0$ or $\pi$ phase shift on the pulse, where the choice of the phase shift determines the detector representing $\vert + \rangle$.
The two pulses are recombined into one in the 45$^\circ$ PBS.
After exiting the Sagnac interferometer, the pulse is split by the PBS.
Two channels of a SNSPD, $D_H$ and $D_V$, are utilized to detect the signals that leave the circulator and the PBS, respectively.
When the insertion loss of the circulator (1.05 dB) is considered, the detection efficiency is approximately 39 $\%$,

The dark count rates of the two SNSPD channels are 24 cps and 5 cps, respectively, and the dead time is 50 ns.  The detection efficiency of the SNSPD is time-independent.
Thus, it is immune to the time-shift attack~\cite{zhao2008quantum}. In our data analysis, all detection events from the entire time period are used in phase error rate estimation instead of using the preset time window. This enables our experimental system to resist dead time attacks~\cite{weier2011quantum} and afterpulse attacks~\cite{afterpulse}. Note that this strategy is effective for both SNSPD and avalanche photodiodes.

The randomness source can theoretically be offered by any other party.
To demonstrate the ability to generate random numbers with our protocol, we desire $\vert H \rangle$ pulses to achieve the best generation performance according to the design of our detection equipment.
Thus, we use the uncharacterized light emitted by a 14-pin butterfly laser diode with a homemade driving circuit and pump it into the slow axis of the polarization maintaining fiber.
The laser is triggered by the arbitrary waveform generator and emits pulses with a 5 MHz repetition rate.
The best scenario is when the output state of the source is $\vert H\rangle$.
If the output state is another polarization state, it affects only the generated randomness per pulse. The security analysis provided here universally fits the unknown input state.

\begin{figure}[t]
\centering
\includegraphics[width=8.6cm]{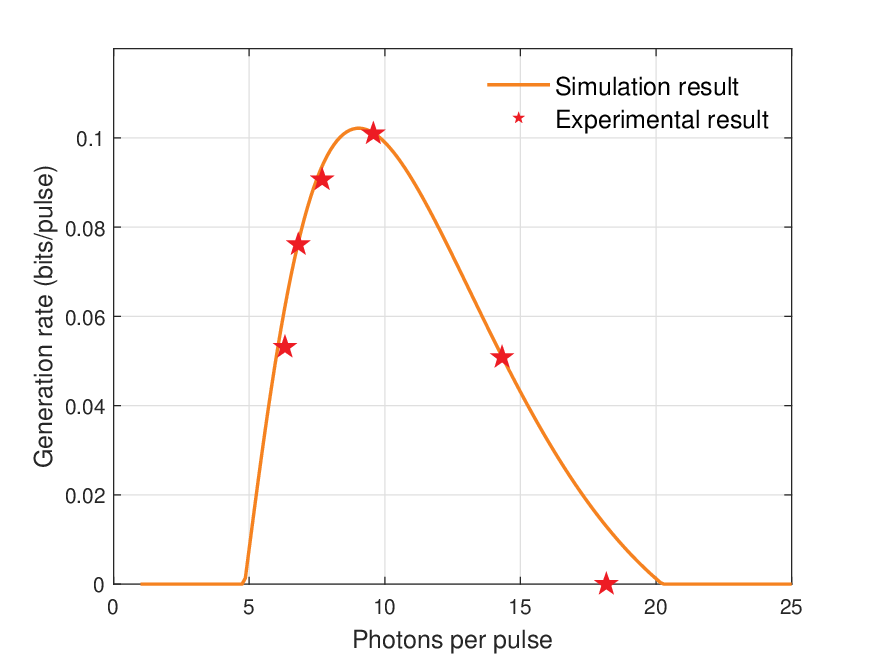}
\caption{Random number generation rates in experiments. The orange line represents the simulation results, and red stars represent the experimental results. In the experiment, the random number generation rate is 0.101 when the intensity is $\mu = 9.6$. With 5 MHz system repetition, we accumulate data for approximately 200 seconds at each point, corresponding to a data size of $10^9$, and a random number generation speed of 505 kbps is achieved.}
\label{fig_intensity_R}
\end{figure}

The intensity of the pulse influences the type of click event and thus affects the generation rate.
In Fig.~\ref{fig_intensity_R}, the abscissa represents the light intensity, and the ordinate represents the generation rate.
The orange line represents the simulation results, and the red pentacles are the experimental results.
In Eq.~(\ref{len}), the value of $q$ should be calibrated.
According to the entropic uncertainty relation, $q=-\log_2{\max_{x,z}{\vert\langle x\vert z \rangle\vert^2}}$ is the incompatibility between two measurement bases.
To realize the calibration, we first modify the light until the ratio of photon counts between the two detection channels is above 24 dB in the $\mathbb{X}$ basis.
This means that the light is approximately a perfect eigenstate of $\mathbb{X}$.
Subsequently, we measure the light in the $\mathbb{Z}$ basis and obtain the ratio of photon counts between the two detection channels. By comparing the single-click events in the two detection channels, the value of $q$ is calibrated to $q=0.954$ in our detection equipment.

The unbalanced detection efficiency should be taken into account~\cite{wei2017trustworthiness}.
Its impact is introduced as a coefficient of the generation rate~\cite{fung2009security,ma2020practical}.
This coefficient is $\eta_{e}=2\min\{(\eta_0,\eta_1)\}/(\eta_0+\eta_1)$, which depends on the efficiencies of the two detection channels.
The final random number extracted is
\begin{equation}\label{len2}
\begin{aligned}  
\ell & \geqslant \eta_{e}\left(N_{z}^{s}[q-h_d(\bar{\phi}_z)]-2\log_2\frac{3}{2\varepsilon_{\rm sec}}\right)-n_{seeds},
\end{aligned}
\end{equation}
The detection efficiencies are $49\%$ and $39\%$, respectively, in calibration. After taking the insert loss circulator (1.05 dB) into account, $\eta_{e}$ is calculated to be $0.9932$.
The detailed experimental results are shown in ''Supplemental Document''.
For each data point, we collect approximately 200 seconds of data, corresponding to a data size of $10^9$. When calculating the error rate in the $\mathbb{X}$ basis, no-click events, single-click events on the incorrect detector, and double-click events are all taken into account.
At the optimal point, the intensity of the pulses before entering the detector is 9.3 photons per pulse, and the random number generation rate is 0.101.

\begin{figure}[t]
\centering
\includegraphics[width=8.6cm]{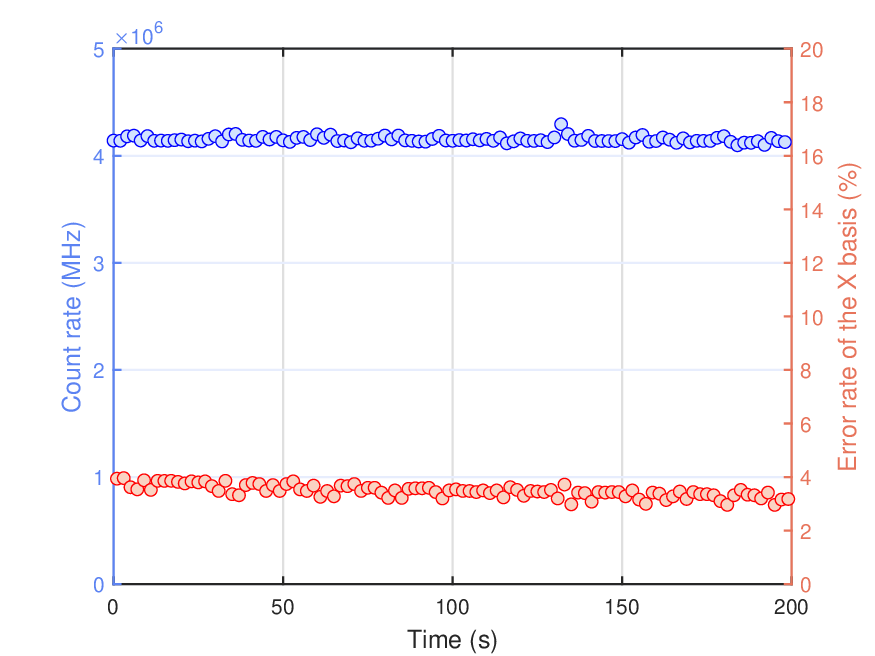}
\caption{The detector count rate and error rate over 200 seconds. We use data for a 10 dB channel loss with optimal intensity. Each dot corresponds to the data acquired over two seconds. During testing, the count rate is always approximately 4.15 MHz. No-click events, incorrect detector click events and double-click events are all treated as errors in the $\mathbb{X}$ basis, and the error rate of the $\mathbb{X}$ basis is always less than 4$\%$, which shows the stability of our experimental system.}
\label{fig_error_time}
\end{figure}

\begin{figure}[t]
\centering
\includegraphics[width=8.6cm]{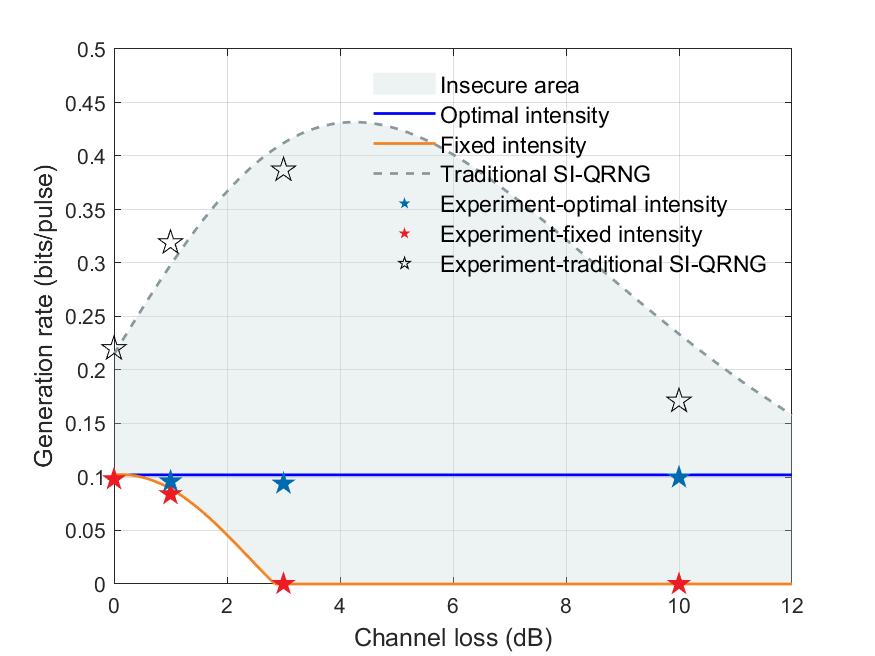}
\caption{Relation between the random number generation rates and the channel loss. The blue line is the generation rate when the intensity is always optimal, and the red line is the generation rate when the intensity has a fixed value of $\mu = 9.3$. The gray dashed line is the generation rate of the traditional SI-QRNG protocol with no-click events discarded provided in ''Supplemental Document''. Stars in different colors represent the experimental results. Discarding no-click events leads to security risks under bright illumination, as shown in the gray filled area.}
\label{fig_loss_R}
\end{figure}

We also consider the influence of the channel loss on random number generation.
The loss reduces the light intensity reaching the detector, thereby decreasing the generation rate when the source produces pulses with the optimal intensity.
Fortunately, our experimental setup enables us to compensate for the channel loss by increasing the intensity of the source. First, to show the stability of the experimental system, the detector count rate and error rate versus time are presented in Fig.~\ref{fig_error_time}. We use data collected with a 10 dB channel loss with optimum intensity. During the 200-second test time, the count rate is always approximately 4.15 MHz (corresponding to $\mu$ = 9.17 before entering the detector), and the error rate of the $\mathbb{X}$ basis is approximately $3.5 \%$ (including both no-click and double-click events). We then experimentally show the relation between the loss and the generation rate under a fixed intensity and variable intensity, as shown in Fig.~\ref{fig_loss_R}.
Our protocol is compared with the traditional source-independent scheme in ''Supplemental Document''.
The difference between the two schemes is whether no-click events are treated as valid events.
At each channel loss, the data obtained with fixed intensity are analyzed to calculate the key rate for both our protocol and the traditional source-independent protocol.
In the case of high channel loss, the generation rate of our protocol is zero due to the
high error rate that results from no-click events. Meanwhile, the traditional source-independent protocol can still generate random numbers, and the generation rate can be as high as 0.387 (corresponding to 1.94 Mbps) when the channel loss is 3 dB. This difference in the generation rate under the same loss reflects the maximal security vulnerabilities caused by tailored detector blinding attacks.  By increasing the intensity of the source to compensate for the channel loss, the generation rate can be maintained at the optimal level.

\begin{table}[t]
\renewcommand\arraystretch{1.2}
\centering
\caption{The results of the NIST statistical test. To pass the test, two criteria should be satisfied. If we set the significance level at $\alpha = 0.01 $, the minimum proportion of sequences that pass a test item is approximately 0.96, except for the random excursion (variant) test, for which the required proportion is 0.952. In addition, the uniformity of the P-value distribution $P$-$value_{~T}$ should be greater than 0.0001. When a test item yields multiple results, the smallest value is reported.}\label{tab1}
\begin{tabular}[b]{lccc}
\hline
\hline
Test name  & $P$-$value_{~T}$ & Proportion & Result \\
\hline
Frequency	&	0.1538	&	1.00	&	Pass	\\
BlockFrequency	&	0.9357	&	0.98	&	Pass	\\
CumulativeSums	&	0.2023	&	1.00	&	Pass	\\
Runs	&	0.1816	&	1.00	&	Pass	\\
LongestRun	&	0.1626	&	0.99	&	Pass	\\
Rank	&	0.9781	&	0.98	&	Pass	\\
FFT	&	0.5955	&	0.98	&	Pass	\\
NonOverlappingTemplate	&	0.0118	&	0.99	&	Pass	\\
OverlappingTemplate	&	0.4559	&	1.00	&	Pass	\\
Universal	&	0.2493	&	0.99	&	Pass	\\
ApproximateEntropy	&	0.4559	&	1.00	&	Pass	\\
RandomExcursions	&	0.0106	&	0.984	&	Pass	\\
RandomExcursionsVariant	&	0.0022	&	0.984	&	Pass	\\
Serial	&	0.8165	&	0.99	&	Pass	\\
LinearComplexity	&	0.5749	&	0.97	&	Pass	\\
\hline
\hline
\end{tabular}
\end{table}

To further verify the quality of the final output random numbers, we apply the standard NIST statistical tests~\cite{rukhin2001statistical}. After collecting data for approximately 200 seconds, a total of $1.02 \times 10^9$ pulses have been sent, and the key rate is 0.103.
After privacy amplification, the final random number of length $1.05\times 10^8$ is divided into 100 bitstreams, and fifteen statistical tests are implemented. As shown in Table~\ref{tab1}, the random numbers generated in the experiment pass all NIST statistical tests.

\section{Discussion}
In conclusion, we have proposed an SI-QRNG type protocol that can resist the tailored detector blinding attack.
By exploiting the uncertainty relation for smooth entropy, our protocol can be easily extended to high-dimensional measurement cases with composable security against coherent attacks under the finite-key effect.

In our experiments, the detection loss and the channel loss can be compensated by improving the emission intensity.
Using a 5 MHz experimental system, we achieve a quantum random number generation speed of over 500 kbps.
By increasing the saturation count rate of the detectors to GHz~\cite{zhang201916}, the QRNG generation rate can be enhanced to more than 100 Mbps.
Through simple experimental tricks, our experimental implementation suppresses most well-known attacks on detector components, realizing an extremely high security level approaching device-independence.
Note that our theoretical framework and experimental scheme are a general solution to tailored detector blinding attacks, therefore it is applicable to both SNSPD and avalanche photodiodes.

Here we briefly discuss why detector blinding attacks need to be studied.
This kind of attack was first proposed in the quantum key distribution tasks. There are several experimental countermeasures against detector blinding attacks for quantum key distribution tasks.
A common solution is to install a beam splitter before signals enter the detection equipment to monitor the light intensity~\cite{BI}.
Crafty adversaries can instead send instantaneous bright trigger light, which blinds the detector without disrupting the monitor~\cite{wiechers2011after,lydersen2011superlinear}.
Other solutions~\cite{Solu3,Solu4,Solu6,Gras2021Coun}, such as randomly changing the attenuation in front of the detector and analyzing the corresponding detection events and errors, also increase the difficulty of experimental operation~\cite{wu2020robust}.
While experimental solutions attempt to judge whether the generator is under attack by designing a more sophisticated system, further advanced attacks from Eve usually cannot be avoided~\cite{sajeed2016insecurity,wu2020robust}.
Finally, they propose the measurement-device-independent scheme that only trust the sources, and the whole detection component is handed over to an untrusted third party.
However, for QRNG, trusting detection component is more reasonable, because there is only one user.
SIQRNG protocols can select sources from local materials, which is more practical.
In this case, the detector blinding attack needs to be carefully considered. Note that recent work has also considered this attack~\cite{Lin2022certified}.

Bell tests~\cite{brunner2014bell} indicate the importance of the fair sampling assumption.
Once the assumption is not established, it is necessary to carefully handle all measurement outcomes, especially no-click events.
comprehensive consideration is an important feature that distinguishes our protocol from others.
Furthermore, using the probability correlation of click events, only part of the no-click events will be count into final results in Bell tests.
Accordingly, our protocol has the opportunity to reduce the impact of no-click events on the error rate through a certain probability correlation to improve the generation rate.

It is worth noting that we are not device-independent protocols.
Thus it is impossible to defend against all attacks on detectors.
Here we solved the tailored detector blinding attack that controls random numbers by directly sending the state corresponding to the desired outcome.
There may also be more complex attacks.
For instance, Eve can carefully analyze detection thresholds of two detectors to generate a superposition state of $\mathbb{Z}$ for attacks.
We can avoid this problem through polarization-maintaining fiber in our experiment.
In addition, when two detectors have different detection thresholds, Eve can only make the detector with small threshold click.
These are all worthy of further consideration

Finally, a passive-basis-choice approach may also help realize random generation.
We apply an active basis choice in our protocol, which consumes a considerable amount of random numbers.
For this reason, our protocol is a random expansion rather than an absolutely random generation.
Passive-basis-choice can avoid this kind of consumption and enables real random extraction through further discarding the double-basis clicks.
To maintain security, some assumptions must be introduced.
It is also worth investigating whether these assumptions are reasonable.

\section*{Funding}
Natural Science Foundation of Jiangsu Province (No. BK20211145); Fundamental Research Funds for the Central Universities (No. 020414380182); Key Research and Development Program of Nanjing Jiangbei New Aera (No. ZDYD20210101); Program for Innovative Talents and Entrepreneurs in Jiangsu (No. JSSCRC2021484).

\section*{Acknowledgments}
We thank P. Liu, X.-Y. Cao and C.-X. Weng for their valuable discussions.


\appendix
\begin{widetext}
\section*{Supplemental document}
This document provides a source-independent protocol that suffers from the tailored detector blinding attacks.
This protocol is the closest to our protocol and the most commonly used protocol without detector blinding attacks.
We compared it with our protocol in the main text, thus offering the simulation formulas of this protocol here.
The statistics of the experimental data are given here for the need of most readers to understand the data analysis.

\section{SI-QRNG protocol treating no-click events as vacua}

In the main text, we provide our protocol that resists tailored detector blinding attacks by treating no-click events as error events.
Previous SI-QRNG protocols did not consider detector blinding attacks, so they used squashing models to mark the no-click event as a vacuum.
Since the vacuum cannot carry information and will not reveal information, they discard no-click events without hesitation.
We compare the random number generation rate of our protocol and this kind of protocol.
The rate difference implies the security risks caused by the detector blinding attack.
To avoid the influence of different squashing model choices and the inconsistency of security parameters under different security frameworks, we use the following protocol as the representative of previous SI-QRNGs.

\subsection{Protocol description}
\noindent\emph{State preparation.} Alice prepares signals that allow the detector representing $\left| + \right\rangle$ to fire.
She sends $N$ signals and does not trust the quantum states she prepared.
Virtual party Bob guesses that the outcomes of measuring signals in $\mathbb{X}$ basis are always $\left| + \right\rangle$.

\noindent\emph{Two-dimensional measurement.} Alice randomly measures signals in basis $\mathbb{X}$ or $\mathbb{Z}$ with probability $p_x$ and $p_z=1-p_x$, respectively. The no-click events are discarded.

\noindent\emph{Post-processing.} Alice only deals with the click events.
In the $\mathbb{X}$ basis, the click events can be divided in two parts: $n^c_x$ and $n^e_x$.
$n^c_x$ is the number of correct events that only the $\left| + \right\rangle$ detector fires.
Other events, including double-click events and single-click events on the $\left| - \right\rangle$ detector, are considered as error results and counted into $n^e_x$.
The error rate is $e_x = n^e_x/n_x$, where $n_x$ is the number of click events in $\mathbb{X}$.
In $\mathbb{Z}$ basis, we care only about the single click results $n^s_z$ of all $n_z$ click events.
According to the uncertainty relation of entropy, the length of secret random numbers with $\varepsilon_{\rm sec}$-secrecy is given by
\begin{equation}\label{len}
\begin{aligned}  
\ell & \geqslant n_{z}^{s}[q-h_d(\bar{\phi}_z)]-2\log_2\frac{3}{2\varepsilon_{\rm sec}}-n_{seeds},
\end{aligned}
\end{equation}
where $\bar{\phi_z}=\frac{(e_x+\gamma(n_z,n_x,e_x,\epsilon_{rand}))\times n_z}{n^s_z}$ is the upper bound of the phase error rate in $\mathbb{Z}$ basis, $h_d$ is the d-dimension Shannon entropy, and q is the incompatibility between two measurement bases.
The active basis choice consumes some random seeds $n_{seeds}=N p_x \log_2 N$.

\subsection{Simulation}
In simulation, we assume that the states are coherent states.
In the $\mathbb{X}$ basis, the gain of click events is $Q_x = 1- (1-p_d)^{2}e^{-\mu\eta}$, where $p_d$ is the dark count, $\mu$ is the intensity of light and $\eta$ is the transmission efficiency.
The number of click events is $n_x = N p_x Q_x$.
The gain of error events is $Q_{x}^{e} = p_d + e_d(Q_x-p_d)$, where $e_d$ is the misalignment error.
The error rate is $e_x = Q_{x}^{e}/Q_x$.

In the $\mathbb{Z}$ basis, the gain of click events is also $Q_z = 1- (1-p_d)^{2}e^{-\mu\eta}$.
The number of click events is $n_z = N p_z Q_z$.
The gain of all single-click events is $Q_{z}^{s} =2(1-p_d)e^{-\mu\eta/2}-2(1-p_d)^{2}e^{-\mu\eta}$.
The number of single-click events is $n^s_z=N p_z Q_{z}^{s}$.

\section{Detailed experimental results}
In Tab.~\ref{table_expresult_u_R},~\ref{table_expresult_loss_R},~\ref{table_expresult_loss_R_optimal}, ~\ref{table_expresult_loss_R_supp} and \ref{table_expresult_finalkey} we report the detailed results obtained in the experiment.  
In all the tables, we report the random number generation rate $R:=\ell / N $, the number of time windows that Alice choose to measure in Z(X) basis $N_{Z(X)}$, the total number of click events in detector $D_H$ ($D_V$) in the Z basis, the number of single click events in detector $D_H$ ($D_V$) in the Z basis, the error rate in the X basis $e_x$ and the upper bound of phase error rate in the Z basis $\overline{\phi}_z$.

\begin{table}[h]
\begin{tabular}{c|cccccc} 
\hline\hline 
  \text{Intensity $\mu$} & 6.32 & 6.8 & 7.7 & 9.6 & 14.3& 18.2  \\
\hline
$R$
&0.0531 
&0.0761 
&0.0907 
&0.1010 
&0.0509 
&0
\\
$N_Z$ & 1064540311 & 1059470000 & 999500000  & 1009708496 & 99549664 & 999511985 \\
$N_X$ & 532592 &530089 & 500049&505161& 500078& 500063 \\
Total $D_H$ click in the Z basis 
&732490456
&795307879
&792392154
&860099255
&939175515
&968019162
\\
Total $D_V$  click in the Z basis 
&722825567
&750713996
&748852702
&835145639
&919880879
&954610805
\\
Single $D_H$ click in the Z basis &235037676
&232246534
&198791921
&148663349
&74850323
&43486700\\
Single $D_V$  click in the Z basis &225372787
&187652651
&155252469
&123709733
&55555687
&30078343
\\
$e_x$
&10.11\%
&7.69\%
&5.74\%
&2.99\%
&1.10\%
&2.19\%
\\
$\overline{\phi}_z$   
&23.95\%
&19.96\%
&16.76\%
&11.65\%
&9.17\%
&31.49\%
\\
\hline
\end{tabular}
\caption{Experiment results obtained for various pulse intensities.}
\label{table_expresult_u_R}
\end{table}

\begin{table}[h]
\begin{tabular}{c|cccccc} 
\hline\hline 
  \text{Channel loss (dB)} & 0 & 1 & 3 & 10  \\
\hline
$R$
&0.0979 
&0.0839
&0
&0
\\
$N_Z$ 
&1009530770
&1149424997
&999592872
&1229842581
\\
$N_X$ 
&505067
&575054
&500083
&615245
\\
Total $D_H$ click in the Z basis 
&826646616
&844496091
&576423481
&143298859
\\
Total $D_V$ click in the Z basis 
&849768774
&861501605
&566073702
&136472595
\\
Single $D_H$ click in the Z basis 
&130668741
&211532227
&249937412
&127373092
\\
Single $D_V$ click in the Z basis 
&153790899
&228537741
&239587633
&120546828
\\
$e_x$
&3.42\%
&6.91\%
&19.64\%
&78.55\%
\\
$\overline{\phi}_z$   
&12.68\%
&18.57\%
&40.78\%
&50\%
\\
\hline
\end{tabular}
\caption{Experiment results obtained for various channel losses with a fixed intensity.}
\label{table_expresult_loss_R}
\end{table}

\begin{table}[h]
\begin{tabular}{c|cccccc} 
\hline\hline 
  \text{Channel loss (dB)} & 1 & 3 & 10  \\
\hline
$R$
&0.0959
&0.0938 
&0.0995
\\
$N_Z$ 
&999980222
&1009494998
&1009882327
\\
$N_X$ 
&500292
&505052
&505248
\\
Total $D_H$ click in the Z basis 
&850571844
&822977603
&828712721
\\
Total $D_V$ click in the Z basis 
&835192945
&851910282
&842746871
\\
Single $D_H$ click in the Z basis 
&140183635
&128031450
&137373945
\\
Single $D_V$ click in the Z basis 
&124804736
&156964129
&151408095
\\
$e_x$
&3.04\%
&3.58\%
&3.47\%
\\
$\overline{\phi}_z$   
&12.05\%
&13.24\%
&12.69\%
\\
\hline
\end{tabular}
\caption{Experiment results obtained for various channel losses with the optimal intensity.}
\label{table_expresult_loss_R_optimal}
\end{table}

\begin{table}[h]
\begin{tabular}{c|cccccc} 
\hline\hline
\text{Channel loss (dB)}& 0 & 1 & 3 & 10  \\
\hline
$R$
&0.220
&0.319
&0.387
&0.171
\\
$N_Z$ 
&1009530770
&1149424997
&999592872
&1229842581
\\
$N_X$ 
&505067
&575054
&500083
&615245
\\
Total $D_H$ click in the Z basis 
&826646616
&844496091
&576423481
&143298859
\\
Total $D_V$ click in the Z basis 
&849768774
&861501605
&566073702
&136472595
\\
Single $D_H$ click in the Z basis 
&130668741
&211532227
&249937412
&127373092
\\
Single $D_V$ click in the Z basis 
&153790899
&228537741
&239587633
&120546828
\\
$e_x$
&0.41\% 
&0.36\%
&1.01\%
&0.20\%
\\
$\overline{\phi}_z$   
&1.62\%
&1.02\%
&1.86\%
&0.32\%
\\
\hline
\end{tabular}
\caption{Experiment results obtained for various channel losses with a fixed intensity, analysed for the protocol in supplementary.}
\label{table_expresult_loss_R_supp}
\end{table}

\begin{table}[h]
\begin{tabular}{c|cccccc}  
\hline \hline
$R$
&0.103
\\
$N_Z$ 
&1015527087
\\
$N_X$ 
&508128
\\
Total $D_H$ click in the Z basis 
&854697340
\\
Total $D_V$ click in the Z basis 
&824303459
\\
Single $D_H$ click in the Z basis 
&160570766
\\
Single $D_V$ click in the Z basis 
&130176885
\\
$e_x$
&3.34\%
\\
$\overline{\phi}_z$   
&12.2\%
\\
\hline
\end{tabular}
\caption{Experimental results used for final random number generation.}
\label{table_expresult_finalkey}
\end{table}
\end{widetext}


\begin{thebibliography}{67}
\newcommand{\enquote}[1]{``#1''}

\bibitem{pserand}
F.~James, \enquote{A review of pseudorandom number generators,} {\bibinfo {journal} {Comput. Phys. Commun.}} \textbf{60}, 329--344 (1990).

\bibitem{pserand3}
P.~L'Ecuyer, \enquote{Random number generation,} in \emph{Handbook of
  Computational Statistics: Concepts and Methods,}  (Springer, 2012, pp. 35--71).

\bibitem{truerand1}
V.~Fischer and M.~Drutarovsk{\`y}, \enquote{True random number generator
  embedded in reconfigurable hardware,} in \emph{Cryptographic Hardware and
  Embedded Systems - CHES 2002,}  (Springer, 2003, pp. 415--430).

\bibitem{truerand2}
I.~Reidler, Y.~Aviad, M.~Rosenbluh, and I.~Kanter, \enquote{Ultrahigh-speed
  random number generation based on a chaotic semiconductor laser,}
  {\bibinfo {journal} {Phys. Rev. Lett.}} \textbf{103}, 024102 (2009).

\bibitem{QRNG1}
X.~Ma, X.~Yuan, Z.~Cao, B.~Qi, and Z.~Zhang, \enquote{Quantum random number
  generation,} {\bibinfo {journal} {npj Quantum Inf.}} \textbf{2}, 16021 (2016).

\bibitem{QRNG2}
M.~Herrero-Collantes and J.~C. Garcia-Escartin, \enquote{Quantum random number
  generators,} {\bibinfo {journal} {Rev. Mod. Phys.}} \textbf{89}, 015004
  (2017).

\bibitem{app1}
J.~Liu, Y.~Qi, Z.~Y. Meng, and L.~Fu, \enquote{Self-learning monte carlo method,} {\bibinfo {journal} {Phys. Rev. B}} \textbf{95}, 041101 (2017).

\bibitem{app2}
N.~Masuda, M.~A. Porter, and R.~Lambiotte, \enquote{Random walks and diffusion
  on networks,} {\bibinfo {journal} {Phys. Rep.}} \textbf{716-717}, 1--58
  (2017).

\bibitem{app4}
M.-O. Renou, D.~Trillo, M.~Weilenmann, T.~P. Le, A.~Tavakoli, N.~Gisin,
  A.~Ac{\'\i}n, and M.~Navascu{\'e}s, \enquote{Quantum theory based on real
  numbers can be experimentally falsified,} {\bibinfo {journal} {Nature}}
  \textbf{600}, 625--629 (2021).

\bibitem{yin2016practical}
H.-L. Yin, Y.~Fu, and Z.-B. Chen, \enquote{Practical quantum digital
  signature,} {\bibinfo {journal} {Phys. Rev. A}} \textbf{93}, 032316 (2016).

\bibitem{app3}
P.~Alikhani, N.~Brunner, C.~Cr{\'e}peau, S.~Designolle, R.~Houlmann, W.~Shi,
  N.~Yang, and H.~Zbinden, \enquote{Experimental relativistic zero-knowledge
  proofs,} {\bibinfo {journal} {Nature}} \textbf{599}, 47--50 (2021).

\bibitem{fu2015long}
Y.~Fu, H.-L. Yin, T.-Y. Chen, and Z.-B. Chen, \enquote{Long-distance
  measurement-device-independent multiparty quantum communication,}
  {\bibinfo {journal} {Phys. Rev. Lett.}} \textbf{114}, 090501 (2015).

\bibitem{app5}
S.~Pirandola, U.~L. Andersen, L.~Banchi, M.~Berta, D.~Bunandar, R.~Colbeck, D.~Englund, T.~Gehring, C.~Lupo, C.~Ottaviani, J. L. Pereira, M. Razavi, J. Shamsul Shaari, M. Tomamichel, V. C. Usenko, G. Vallone, P. Villoresi, and P. Wallden, \enquote{Advances in quantum cryptography,} {\bibinfo {journal} {Adv. Opt. Photon.}} \textbf{12}, 1012--1236 (2020).

\bibitem{liu2021homodyne}
W.-B. Liu, C.-L. Li, Y.-M. Xie, C.-X. Weng, J.~Gu, X.-Y. Cao, Y.-S. Lu, B.-H.
  Li, H.-L. Yin, and Z.-B. Chen, \enquote{Homodyne detection quadrature phase
  shift keying continuous-variable quantum key distribution with high excess
  noise tolerance,} {\bibinfo {journal} {PRX Quantum}} \textbf{2}, 040334
  (2021).

\bibitem{DI1}
S.~Pironio, A.~Ac{\'\i}n, S.~Massar, A.~B. de~La~Giroday, D.~N. Matsukevich,
  P.~Maunz, S.~Olmschenk, D.~Hayes, L.~Luo, T.~A. Manning, and C. Monroe,
  \enquote{Random numbers certified by bell's theorem,}
  {\bibinfo {journal} {Nature}} \textbf{464}, 1021--1024 (2010).

\bibitem{DI2}
Y.~Liu, Q.~Zhao, M.-H. Li, J.-Y. Guan, Y.~Zhang, B.~Bai, W.~Zhang, W.-Z. Liu, C.~Wu, X.~Yuan, H.~Li, W. J. Munro, Z. Wang, L. You, J. Zhang, X. Ma, J. Fan, Q. Zhang, and J.-W. Pan, \enquote{Device-independent quantum
  random-number generation,} {\bibinfo {journal} {Nature}} \textbf{562},
  548--551 (2018).

\bibitem{DI3}
Y.~Zhang, L.~K. Shalm, J.~C. Bienfang, M.~J. Stevens, M.~D. Mazurek, S.~W. Nam, C.~Abell\'an, W.~Amaya, M.~W. Mitchell, H.~Fu, C.~A. Miller, A.~Mink, and E.~Knill, \enquote{Experimental low-latency device-independent quantum
  randomness,} {\bibinfo {journal} {Phys. Rev. Lett.}} \textbf{124}, 010505
  (2020).

\bibitem{DI4}
P.~Bierhorst, E.~Knill, S.~Glancy, Y.~Zhang, A.~Mink, S.~Jordan, A.~Rommal,
  Y.-K. Liu, B.~Christensen, S.~W. Nam, M. J. Stevens, and L. K. Shalm, \enquote{Experimentally
  generated randomness certified by the impossibility of superluminal signals,}
  {\bibinfo {journal} {Nature}} \textbf{556}, 223--226 (2018).

\bibitem{DI5}
M.-H. Li, X.~Zhang, W.-Z. Liu, S.-R. Zhao, B.~Bai, Y.~Liu, Q.~Zhao, Y.~Peng,
  J.~Zhang, Y.~Zhang, W.~J. Munro, X.~Ma, Q.~Zhang, J.~Fan, and J.-W. Pan,
  \enquote{Experimental realization of device-independent quantum randomness
  expansion,} {\bibinfo {journal} {Phys. Rev. Lett.}} \textbf{126}, 050503
  (2021).

\bibitem{DI6}
L.~K. Shalm, Y.~Zhang, J.~C. Bienfang, C.~Schlager, M.~J. Stevens, M.~D.
  Mazurek, C.~Abell{\'a}n, W.~Amaya, M.~W. Mitchell, M.~A. Alhejji, H. Fu, J. Ornstein, R. P. Mirin, S. W. Nam, and E. Knill, \enquote{Device-independent randomness expansion with
  entangled photons,} {\bibinfo {journal} {Nat. Phys.}} \textbf{17}, 452--456
  (2021).

\bibitem{DI7}
W.-Z. Liu, M.-H. Li, S.~Ragy, S.-R. Zhao, B.~Bai, Y.~Liu, P.~J. Brown,
  J.~Zhang, R.~Colbeck, J.~Fan, Q. Zhang, and J.-W. Pan, \enquote{Device-independent
  randomness expansion against quantum side information,}
  {\bibinfo {journal} {Nat. Phys.}} \textbf{17}, 448--451 (2021).

\bibitem{lowlatency}
Y.~Zhang, H.-P. Lo, A.~Mink, T.~Ikuta, T.~Honjo, H.~Takesue, and W.~J. Munro,
  \enquote{A simple low-latency real-time certifiable quantum random number
  generator,} {\bibinfo {journal} {Nat. Commun.}} \textbf{12}, 1056 (2021).

\bibitem{SDIF}
A.~Tavakoli, \enquote{Semi-device-independent framework based on restricted
  distrust in prepare-and-measure experiments,} {\bibinfo {journal} {Phys.
  Rev. Lett.}} \textbf{126}, 210503 (2021).

\bibitem{MDIF}
Y.-Q. Nie, J.-Y. Guan, H.~Zhou, Q.~Zhang, X.~Ma, J.~Zhang, and J.-W. Pan,
  \enquote{Experimental measurement-device-independent quantum random-number
  generation,} {\bibinfo {journal} {Phys. Rev. A}} \textbf{94}, 060301
  (2016).

\bibitem{QSI}
T.~Gehring, C.~Lupo, A.~Kordts, D.~Solar~Nikolic, N.~Jain, T.~Rydberg, T.~B.
  Pedersen, S.~Pirandola, and U.~L. Andersen, \enquote{Homodyne-based quantum
  random number generator at 2.9 gbps secure against quantum side-information,}
  {\bibinfo {journal} {Nat. Commun.}} \textbf{12}, 605 (2021).

\bibitem{SI1}
Z.~Cao, H.~Zhou, X.~Yuan, and X.~Ma, \enquote{Source-independent quantum random
  number generation,} {\bibinfo {journal} {Phys. Rev. X}} \textbf{6}, 011020
  (2016).

\bibitem{SDI1}
D.~G. Marangon, G.~Vallone, and P.~Villoresi,
  \enquote{Source-device-independent ultrafast quantum random number
  generation,} {\bibinfo {journal} {Phys. Rev. Lett.}} \textbf{118}, 060503
  (2017).

\bibitem{SDI2}
M.~Avesani, D.~G. Marangon, G.~Vallone, and P.~Villoresi,
  \enquote{Source-device-independent heterodyne-based quantum random number
  generator at 17 gbps,} {\bibinfo {journal} {Nat. Commun.}} \textbf{9}, 5365
  (2018).

\bibitem{SDI3}
D.~Drahi, N.~Walk, M.~J. Hoban, A.~K. Fedorov, R.~Shakhovoy, A.~Feimov, Y.~Kurochkin, W.~S. Kolthammer, J.~Nunn, J.~Barrett, and I.~A. Walmsley, \enquote{Certified quantum random numbers from untrusted light,} {\bibinfo {journal} {Phys. Rev. X}} \textbf{10}, 041048 (2020).

\bibitem{cheng2022mutually}
J.~Cheng, J.~Qin, S.~Liang, J.~Li, Z.~Yan, X.~Jia, and K.~Peng,
  \enquote{Mutually testing source-device-independent quantum random number
  generator,} {\bibinfo {journal} {Photon. Res.}} \textbf{10}, 646--652
  (2022).

\bibitem{afterpulse}
X.~Lin, S.~Wang, Z.-Q. Yin, G.-J. Fan-Yuan, R.~Wang, W.~Chen, D.-Y. He,
  Z.~Zhou, G.-C. Guo, and Z.-F. Han, \enquote{Security analysis and improvement
  of source independent quantum random number generators with imperfect
  devices,} {\bibinfo {journal} {npj Quantum Inf.}} \textbf{6}, 100 (2020).

\bibitem{sunlight}
Y.-H. Li, X.~Han, Y.~Cao, X.~Yuan, Z.-P. Li, J.-Y. Guan, J.~Yin, Q.~Zhang,
  X.~Ma, C.-Z. Peng, and J.-W. Pan, \enquote{Quantum random number generation
  with uncharacterized laser and sunlight,} {\bibinfo {journal} {npj Quantum
  Inf.}} \textbf{5}, 97 (2019).

\bibitem{APDattack}
L.~Lydersen, C.~Wiechers, C.~Wittmann, D.~Elser, J.~Skaar, and V.~Makarov,
  \enquote{Hacking commercial quantum cryptography systems by tailored bright
  illumination,} {\bibinfo {journal} {Nat. Photonics}} \textbf{4}, 686--689
  (2010).

\bibitem{wiechers2011after}
C.~Wiechers, L.~Lydersen, C.~Wittmann, D.~Elser, J.~Skaar, C.~Marquardt,
  V.~Makarov, and G.~Leuchs, \enquote{After-gate attack on a quantum
  cryptosystem,} {\bibinfo {journal} {New J. Phys.}} \textbf{13}, 013043
  (2011).

\bibitem{lydersen2011superlinear}
L.~Lydersen, N.~Jain, C.~Wittmann, {\O}.~Mar{\o}y, J.~Skaar, C.~Marquardt,
  V.~Makarov, and G.~Leuchs, \enquote{Superlinear threshold detectors in
  quantum cryptography,} {\bibinfo {journal} {Phys. Rev. A}} \textbf{84},
  032320 (2011).

\bibitem{app6}
F.~Xu, X.~Ma, Q.~Zhang, H.-K. Lo, and J.-W. Pan, \enquote{Secure quantum key
  distribution with realistic devices,} {\bibinfo {journal} {Rev. Mod.
  Phys.}} \textbf{92}, 025002 (2020).

\bibitem{APDattackfull}
I.~Gerhardt, Q.~Liu, A.~Lamas-Linares, J.~Skaar, C.~Kurtsiefer, and V.~Makarov,
  \enquote{Full-field implementation of a perfect eavesdropper on a quantum
  cryptography system,} {\bibinfo {journal} {Nat. Commun.}} \textbf{2}, 349
  (2011).

\bibitem{nanoattack}
L.~Lydersen, M.~K. Akhlaghi, A.~H. Majedi, J.~Skaar, and V.~Makarov,
  \enquote{Controlling a superconducting nanowire single-photon detector using
  tailored bright illumination,} {\bibinfo {journal} {New J. Phys.}}
  \textbf{13}, 113042 (2011).

\bibitem{brunner2014bell}
N.~Brunner, D.~Cavalcanti, S.~Pironio, V.~Scarani, and S.~Wehner, \enquote{Bell
  nonlocality,} {\bibinfo {journal} {Rev. Mod. Phys.}} \textbf{86}, 419
  (2014).

\bibitem{Lusquash1}
N.~J. Beaudry, T.~Moroder, and N.~L\"utkenhaus, \enquote{Squashing models for
  optical measurements in quantum communication,} {\bibinfo {journal} {Phys.
  Rev. Lett.}} \textbf{101}, 093601 (2008).

\bibitem{Losquash}
C.-H.~F. Fung, H.~F. Chau, and H.-K. Lo, \enquote{Universal squash model for
  optical communications using linear optics and threshold detectors,}
  {\bibinfo {journal} {Phys. Rev. A}} \textbf{84}, 020303 (2011).

\bibitem{Lusquash2}
O.~Gittsovich, N.~J. Beaudry, V.~Narasimhachar, R.~R. Alvarez, T.~Moroder, and N.~L\"utkenhaus, \enquote{Squashing model for detectors and applications to
  quantum-key-distribution protocols,} {\bibinfo {journal} {Phys. Rev. A}}
  \textbf{89}, 012325 (2014).

\bibitem{Kisquash}
T.~Tsurumaru and K.~Tamaki, \enquote{Security proof for quantum-key-distribution systems with threshold detectors,}
  {\bibinfo {journal} {Phys. Rev. A}} \textbf{78}, 032302 (2008).

\bibitem{Smooth}
M.~Tomamichel and R.~Renner, \enquote{Uncertainty relation for smooth
  entropies,} {\bibinfo {journal} {Phys. Rev. Lett.}} \textbf{106}, 110506
  (2011).

\bibitem{QKDex}
M.~Tomamichel, C.~C.~W. Lim, N.~Gisin, and R.~Renner, \enquote{Tight finite-key
  analysis for quantum cryptography,} {\bibinfo {journal} {Nat. Commun.}}
  \textbf{3}, 634 (2012).

\bibitem{randsamp}
H.-L. Yin, M.-G. Zhou, J.~Gu, Y.-M. Xie, Y.-S. Lu, and Z.-B. Chen,
  \enquote{Tight security bounds for decoy-state quantum key distribution,}
  {\bibinfo {journal} {Sci. Rep.}} \textbf{10}, 14312 (2020).

\bibitem{Shan1}
Y.~Ding, D.~Bacco, K.~Dalgaard, X.~Cai, X.~Zhou, K.~Rottwitt, and L.~K.
  Oxenl{\o}we, \enquote{High-dimensional quantum key distribution based on
  multicore fiber using silicon photonic integrated circuits,}
  {\bibinfo {journal} {npj Quantum Inf.}} \textbf{3}, 25 (2017).

\bibitem{Shan2}
N.~T. Islam, C.~C.~W. Lim, C.~Cahall, J.~Kim, and D.~J. Gauthier,
  \enquote{Provably secure and high-rate quantum key distribution with time-bin
  qudits,} {\bibinfo {journal} {Sci. Adv.}} \textbf{3}, e1701491 (2017).

\bibitem{leftover}
M.~Tomamichel, C.~Schaffner, A.~Smith, and R.~Renner, \enquote{Leftover hashing
  against quantum side information,} {\bibinfo {journal} {IEEE Trans. Inf.
  Theory}} \textbf{57}, 5524--5535 (2011).

\bibitem{li2011attacking}
H.-W. Li, S.~Wang, J.-Z. Huang, W.~Chen, Z.-Q. Yin, F.-Y. Li, Z.~Zhou, D.~Liu,
  Y.~Zhang, G.-C. Guo, W.-S. Bao, and Z.-F. Han, \enquote{Attacking a practical
  quantum-key-distribution system with wavelength-dependent beam-splitter and
  multiwavelength sources,} {\bibinfo {journal} {Phys. Rev. A}} \textbf{84}, 062308 (2011).

\bibitem{pinheiro2018eavesdropping}
P.~V.~P. Pinheiro, P.~Chaiwongkhot, S.~Sajeed, R.~T. Horn, J.-P. Bourgoin,
  T.~Jennewein, N.~L{\"u}tkenhaus, and V.~Makarov, \enquote{Eavesdropping and
  countermeasures for backflash side channel in quantum cryptography,}
  {\bibinfo {journal} {Opt. Express}} \textbf{26}, 21020--21032 (2018).

\bibitem{yin2020experimental}
H.-L. Yin, P.~Liu, W.-W. Dai, Z.-H. Ci, J.~Gu, T.~Gao, Q.-W. Wang, and Z.-Y.
  Shen, \enquote{Experimental composable security decoy-state quantum key
  distribution using time-phase encoding,} {\bibinfo {journal} {Opt. Express}} \textbf{28}, 29479--29485 (2020).

\bibitem{zhao2008quantum}
Y.~Zhao, C.-H.~F. Fung, B.~Qi, C.~Chen, and H.-K. Lo, \enquote{Quantum hacking:
  Experimental demonstration of time-shift attack against practical
  quantum-key-distribution systems,} {\bibinfo {journal} {Phys. Rev. A}}
  \textbf{78}, 042333 (2008).

\bibitem{weier2011quantum}
H.~Weier, H.~Krauss, M.~Rau, M.~F{\"u}rst, S.~Nauerth, and H.~Weinfurter,
  \enquote{Quantum eavesdropping without interception: an attack exploiting the
  dead time of single-photon detectors,} {\bibinfo {journal} {New J. Phys.}}
  \textbf{13}, 073024 (2011).

\bibitem{wei2017trustworthiness}
K.~Wei, H.~Ma, and X.~Yang, \enquote{Trustworthiness of devices in a quantum random number generator based on a symmetric beam splitter,}
  {\bibinfo {journal} {J. Opt. Soc. Am. B}} \textbf{34}, 2185--2189 (2017).

\bibitem{fung2009security}
C.-h.~F. Fung, K.~Tamaki, B.~Qi, H.-K. Lo, and X.~Ma, \enquote{Security proof
  of quantum key distribution with detection efficiency mismatch,}
  {\bibinfo {journal} {Quantum Inf. Comput.}} \textbf{9}, 131 (2009).

\bibitem{ma2020practical}
D.~Ma, Y.~Wang, and K.~Wei, \enquote{Practical source-independent quantum
  random number generation with detector efficiency mismatch,}
  {\bibinfo {journal} {Quantum Inf. Process.}} \textbf{19}, 384 (2020).

\bibitem{rukhin2001statistical}
A.~Rukhin, J.~Soto, J.~Nechvatal, M.~Smid, E.~Barker, S. Leigh, M. Levenson, M. Vangel, D. Banks, N. Heckert, J. Dray, and S. Vo \enquote{A statistical test suite for random and pseudorandom number generators for cryptographic applications,} Special Publication (NIST SP), National Institute of Standards and Technology, Gaithersburg, MD (Accessed February 3, 2023)

\bibitem{zhang201916}
W.~Zhang, J.~Huang, C.~Zhang, L.~You, C.~Lv, L.~Zhang, H.~Li, Z.~Wang, and
  X.~Xie, \enquote{A 16-pixel interleaved superconducting nanowire single-photon detector array with a maximum count rate exceeding 1.5 ghz,}
  {\bibinfo {journal} {IEEE Trans. Appl. Supercond.}} \textbf{29}, 2200204 (2019).

\bibitem{BI}
Z.~Yuan, J.~F. Dynes, and A.~J. Shields, \enquote{Avoiding the blinding attack in qkd,} {\bibinfo {journal} {Nat. Photonics}} \textbf{4}, 800--801 (2010).

\bibitem{Solu3}
Y.-J. Qian, D.-Y. He, S.~Wang, W.~Chen, Z.-Q. Yin, G.-C. Guo, and Z.-F. Han,
  \enquote{Robust countermeasure against detector control attack in a practical quantum key distribution system,} {\bibinfo {journal} {Optica}} \textbf{6}, 1178--1184 (2019).

\bibitem{Solu4}
M.~Fujiwara, T.~Honjo, K.~Shimizu, K.~Tamaki, and M.~Sasaki, \enquote{Characteristics of superconducting single photon detector in dps-qkd system under bright illumination blinding attack,}
  {\bibinfo {journal} {Opt. Express}} \textbf{21}, 6304--6312 (2013).

\bibitem{Solu6}
C.~C.~W. Lim, N.~Walenta, M.~Legr{\'e}, N.~Gisin, and H.~Zbinden,
  \enquote{Random variation of detector efficiency: A countermeasure against
  detector blinding attacks for quantum key distribution,}
  {\bibinfo {journal} {IEEE J. Sel. Top. Quantum Electron.}} \textbf{21},
  192--196 (2015).

\bibitem{Gras2021Coun}
G.~Gras, D.~Rusca, H.~Zbinden, and F.~Bussi\`eres, \enquote{Countermeasure
  against quantum hacking using detection statistics,}
  {\bibinfo {journal} {Phys. Rev. Applied}} \textbf{15}, 034052 (2021).

\bibitem{wu2020robust}
Z.~Wu, A.~Huang, X.~Qiang, J.~Ding, P.~Xu, X.~Fu, and J.~Wu, \enquote{Robust
  countermeasure against detector control attack in a practical quantum key
  distribution system: comment,} {\bibinfo {journal} {Optica}} \textbf{7},
  1391--1393 (2020).

\bibitem{sajeed2016insecurity}
S.~Sajeed, A.~Huang, S.~Sun, F.~Xu, V.~Makarov, and M.~Curty,
  \enquote{Insecurity of detector-device-independent quantum key distribution,}
  {\bibinfo {journal} {Phys. Rev. Lett.}} \textbf{117}, 250505 (2016).

\bibitem{Lin2022certified}
X.~Lin, R.~Wang, S.~Wang, Z.-Q. Yin, W.~Chen, G.-C. Guo, and Z.-F. Han,
  \enquote{Certified randomness from untrusted sources and uncharacterized
  measurements,} {\bibinfo {journal} {Phys. Rev. Lett.}} \textbf{129}, 050506
  (2022).

\end{thebibliography}
\end{document}